\documentclass[twoside]{ilcws08}
\usepackage[latin1]{inputenc}
\usepackage[dvips]{graphicx,epsfig,color,colortbl}
\usepackage{wrapfig,rotating}
\usepackage{amssymb,amsmath,array,url}

\newcommand{\wt}{\widetilde}
\newcommand{\eqn}{equation}

\newcommand{\whizard}{\texttt{WHIZARD}}
\newcommand{\phan}{\phantom{3}}

\begin{document}
\title{
Pinning Down the Invisible Sneutrino at the ILC\footnote{IPPP/09/05, DCPT/09/10, PITHA-09-05, SFB-CPP-09-08}} 
\author{J.~Kalinowski,$^1$ W.~Kilian,$^2$  J.~Reuter,$^3$ T.~Robens,$^4$ and
 K.~Rolbiecki$^5$
\vspace{.3cm}\\
1- Physics Department, University of Warsaw, 00-681 Warsaw, Poland
\vspace{.1cm}\\
2- University of Siegen, Faculty of Physics, D--57068 Siegen, Germany
\vspace{.1cm}\\
3- University of Freiburg, Institute of Physics, D--79104 Freiburg, Germany
\vspace{.1cm}\\
4- RWTH Aachen, Institute for Theoretical Physics E, D--52056 Aachen, Germany
\vspace{.1cm}\\
5- University of Durham, IPPP, Durham, DH1 3LE, UK
}

\maketitle
\begin{abstract}
Using the event generator {\rm \whizard} we study in a realistic ILC environment the prospects of
measuring properties of sneutrinos that decay invisibly into the lightest neutralino and the neutrino.
\end{abstract}

\section{Introduction}
If sneutrinos of the MSSM are
lighter than the lightest chargino and next-to-lightest neutralino, their
decays to charged particles  are  of higher
order and strongly suppressed. Therefore
sneutrinos decay completely invisibly into a neutrino and the
lightest supersymmetric particle (LSP, the lightest neutralino in
the considered case). As a result, they cannot be directly detected in cascade decays at the LHC and a threshold scan at the ILC is
precluded for the very same reason. In such a case, the only possibility to discover
sneutrinos is to select a
well-reconstructible process where the sneutrino is exchanged or shows up as part of a cascade decay, and in which the precise
determination of kinematic distributions gives an access to the
sneutrino mass.
This idea has already been exploited in
Ref.~\cite{Freitas:2005et}, where the lepton-sneutrino decay channel of
charginos produced at the ILC has been investigated.  It was argued that background
effects are sufficiently under control, so a precise mass
determination is possible.
Here we report on results of our study for determining sneutrino properties in
the SPS1a' 
scenario assuming a realistic ILC environment which includes beamstrahlung, initial-state
radiation, a complete account of reducible backgrounds from SM\footnote{An additional SM background due to photon-induced muon pairs, not discussed previously, is included. We thank Graham Wilson for calling this to our attention.} and
SUSY processes, and a complete matrix-element calculation of the
SUSY signal encompassing all irreducible background and
interference contributions~\cite{Kalinowski:2008fk} using the event generator
{\rm \whizard}~\cite{whizard}.

\section{Signal and background}
The sneutrino
mass can be
determined by investigating the chargino decay modes $\wt{\chi}^+_1\to
\wt{\nu}_\ell \ell^+$, $\ell=e,\mu$,
in chargino pair production process at the ILC. In the following we assume that the first and second generation sneutrinos,
$\wt{\nu}_e$ and $\wt{\nu}_\mu$, are mass degenerate.
Large backgrounds from same-flavor lepton and
slepton production can be avoided by selecting events with leptons
of different flavor, i.e.\ one of the charginos decaying to an
electron, the other to a muon. Therefore, we search for (semi-)exclusive
final states:
\begin{\eqn}
\label{eq:mainproc}
  e^+e^- \,\rightarrow\, \wt{\chi}^+_1 \wt{\chi}^-_1\, \rightarrow
\,
  \wt{\nu}_e^*e^-\wt{\nu}_\mu \mu^+\, \rightarrow \,
  e^-\,\mu^+ \,\bar{\nu}_e \,\nu_\mu\,\wt{\chi}^0_1\,\wt{\chi}^0_1\,.
\end{\eqn}
In the SPS1a' scenario  $m_{\wt{\chi}^+_1} = 183.7$~GeV,
$m_{\wt{\nu}_e}=172.5$~GeV and the Born cross section for chargino pair production is predicted to be 173.6 fb at a CM energy of 500 GeV. With a
BR($\wt{\chi}^+_1\to \wt{\nu}_\ell\ell^+ $)=13\%, a sufficient sample of
events can be expected in this channel. The two-body chargino decay
leads to a uniform decay lepton energy spectrum with edges
determined by the chargino and sneutrino masses, which provides a
method for their experimental determination~\cite{Freitas:2005et,Kalinowski:2008fk}. Experimentally, the signature for
the signal process, Eq.~(\ref{eq:mainproc}), is one electron,
one anti-muon and missing energy,\footnote{Equally well, one can
consider $e^+\, \mu^-\,+\,/\!\!\!\!E$ as a final signature or use both
channels.}
\begin{equation} \label{eq:signature}
  e^+ e^- \, \rightarrow \,e^- \mu^+ \,+\, /\!\!\!\!E \; .
\end{equation}

The same final state, Eq.~(\ref{eq:mainproc}), can originate from
other production processes as well as from interference terms, e.g.~multi-peripheral
diagrams, chargino decays into neutralino and $W$, smuon
$\wt{\mu}^+_i \wt{\mu}_j^-$ and selectron $\wt{e}^+_i \wt{e}_j^-$
pair production, and single-resonant SM di-boson production etc.
They will be called collectively as an `irreducible background', in contrast to
the `reducible' SUSY background that includes more neutrinos in the final state. The
latter contribution come from
processes that lead to one or two leptonically decaying $\tau$'s. They
mainly originate from the production of mixed neutralino pairs
$\wt{\chi}^0_i\wt{\chi}^0_j$, stau pairs as well as chargino pairs.
The latter undergo the subsequent decays to stau and LSP or tau
and sneutrino.
\begin{table}[t]
\begin{center}
\begin{tabular}{|c|l|r|r|} \hline
{ Process} &
\mbox{} \hspace{1.5cm} { 500 GeV}
& $\sigma$ [fb], presel.  &  $\sigma^{\text{cut}}$ [fb]\\
\hline\hline
Signal &
$ee \to e \mu\bar\nu_e \nu_\mu
 \tilde{\chi}_1^0\tilde{\chi}_1^0$
 & $3.940(\phan8)$ & $1.639(3)$
\\ \hline\hline
SUSY $\tau$ Bkgd. &
$ee \to \tau\tau
\tilde{\chi}_1^0\tilde{\chi}_1^0 \to
e \mu
\tilde{\chi}_1^0\tilde{\chi}_1^0 4 \nu
$
 & $4.107(\phan 7)$ & $0.978(2)$
\\\hline
SUSY $\tau\nu$ Bkgd. &
$ee \to \tau\tau
\tilde{\chi}_1^0\tilde{\chi}_1^0 2\nu
\to e \mu
\tilde{\chi}_1^0\tilde{\chi}_1^0 6 \nu
$
 & $3.245(10)$ & $0.818(3)$
\\\hline
SUSY $\tau e$ Bkgd. &
$ee \to e \tau \bar\nu_e \nu_\tau
 \tilde{\chi}_1^0\tilde{\chi}_1^0\to e \mu
\tilde{\chi}_1^0\tilde{\chi}_1^0 4 \nu$
 & $3.691(\phan 9)$ & $1.102(8)$
\\\hline
SUSY $\tau \mu$ Bkgd. &
$ee \to \mu \tau \nu_\mu \bar\nu_\tau
 \tilde{\chi}_1^0\tilde{\chi}_1^0 \to e \mu
\tilde{\chi}_1^0\tilde{\chi}_1^0 4 \nu$
 & $2.617(10)$ & $0.966(8)$
\\\hline\hline
SM $WW$ Bkgd. &
$e e \to e \mu 2\nu$
 & $152.42(25)\phan$ & $0.736(2)$
\\\hline
SM $e\tau$ Bkgd. &
$e e \to e \tau 2\nu \to e \mu 4\nu $
 & $26.522(12)$ & $0.317(1)$
\\\hline
SM $\mu\tau$ Bkgd. &
$e e \to \mu \tau 2\nu \to e \mu 4\nu $
 & $15.569(54)$ & $0.174(1)$
\\\hline
SM $\nu$ Bkgd. &
$ e e \to e \mu  4\nu$
 & $0.145(\phan 1)$ &  $0.016(3)$
\\\hline
SM $\tau$ Bkgd. &
$ e e \to \tau \tau \to e \mu 4 \nu$
 & $32.679(98)$ & $<0.001$
\\\hline
SM $\tau\nu$ Bkgd. &
$ e e \to \tau \tau 2\nu \to e \mu
6 \nu$
& $3.852(10)$ & $0.335(9)$
\\\hline\hline
SM $ee\mu\mu$ Bkgd. & $ ee \to ee\mu\mu $
 & $ 522 400(700)\phan\phan\phan $ & $0.105(3) $
\\\hline\hline
SM $\gamma \to \tau$ Bkgd. &
$\gamma^*\gamma^* \to \tau\tau \to e\mu 2\nu$
 & $21392(70)\phan\phan\phan$ & $0.273(2)$
\\\hline
SM $\gamma \to c$ Bkgd. &
$\gamma^*\gamma^* \to c\bar c \to e\mu jj2\nu$
 & $1089 (\phan 4)\phan\phan\phan$ & $<0.001$
\\\hline
SM $\gamma \to W$ Bkgd. &
$\gamma^*\gamma^* \to WW \to e\mu 2\nu$
 & $1.094(\phan 6)$ & $0.079(1)$
\\\hline
SM $\gamma \to \tau\nu_\tau$ Bkgd. &
$\gamma^*\gamma^* \to \tau\tau 2 \nu \to e\mu 8\nu$
 & $0.077(\phan 1)$ & $<0.001$
\\\hline
SM $\gamma\to \ell\tau$ Bkgd. &
$\gamma^*\gamma^* \to (e,\mu) \tau 2\nu \to e \mu 4\nu $
 & $0.404(\phan 2)$ & $0.055(2)$
\\\hline
\end{tabular}
\caption{Cross sections for signal and relevant background processes for
an ILC
  $\sqrt{s}=500$~GeV. ISR and beamstrahlung are always included.
  The `presel.' column includes a $5^\circ$ cut
  for the final electron to cut out collinear regions, and for the $ee\mu\mu$ SM as well as all
  $\gamma$-induced processes a $1^\circ$ cut for particles vanishing
  in the beampipe. The last column shows cross sections after the cuts. In parentheses are the {\rm
    \whizard} integration errors.
\label{tab:500gevxsec}}
\end{center}
\vspace{-7mm}
\end{table}
\noindent The Standard Model background processes leading to the
same signature of different-flavor opposite-sign lepton pairs and missing
energy (carried away by neutrinos) include $WW$ pairs, single
$W$ production and $\tau^+ \tau^-$ pairs. The most severe contribution here is
due to photon-induced processes, especially photon-induced $\mu$- and $\tau$-pair
production. Therefore, the proper simulation of these processes is crucial in
our analysis. As stressed in~\cite{Kalinowski:2008fk}, one has to go beyond the usual
equivalent photon approximation (EPA), since EPA tweaks kinematic
distributions  which results in underestimating the number of background events.
The exact matrix elements for the photon-induced processes need to be generated using quadruple-precision.


\begin{wraptable}{l}{0.5\columnwidth}
\begin{center}
\begin{tabular}{|l|cc|}
\hline  \hspace{1cm} Cut
& lower & upper \\
\hline\hline
  $E(e^-)$, $E(\mu^+)$ & 1 GeV & 40 GeV
\\
  $\theta(e^-)$   &   $15\,^\circ$   &   $155\,^\circ$
\\
  $\theta(\mu^+)$ &   $25\,^\circ$   &   $165\,^\circ$
\\
  $\Delta\phi(e^-,\mu^+)$   &   $-150\,^\circ$   &   $150\,^\circ$
\\
  $p_\perp(e^-)$, $p_\perp(\mu^+)$  & 2 GeV &
\\
  $|\vec{p}_\perp(e^-) + \vec{p}_\perp(\mu^+)|$ & 4 GeV &
\\\hline
\end{tabular}
\caption{\label{tab:cutsall} Cuts used in the analysis for an ILC
with 500 GeV.}
\end{center}
\end{wraptable}
\noindent
As seen in Tab.~\ref{tab:500gevxsec}, the most severe background
(that exceeds the signal by a factor $\geq 10^4$) comes from gamma-induced $\mu$
and $\tau$ pairs which produce  a soft  $e\mu$ system  with small total transverse momentum $p_\perp$ and large angular separation $\Delta \phi$. Therefore, a cut on transverse momentum of leptons and a cut
on azimuthal separation of $e$ and $\mu$ reduces gamma-induced backgrounds by 3 to 5 orders of magnitude.
The other large source of SM background -- $WW$ pair production -- can be reduced by applying an upper cut
on the lepton energy, since most leptons from $W$ decays have energies larger than 40~GeV.
We also require two well-visible leptons in the central part of the detector.
All cuts are summarized in Tab.~\ref{tab:cutsall}. These cuts are less effective in suppressing the
SUSY background processes, since leptons originating from decays of
heavy  SUSY states, with almost degenerate spectrum like SPS1a'~\cite{AguilarSaavedra:2005pw}, are rather soft and angularly uncorrelated.
\begin{figure}[t]
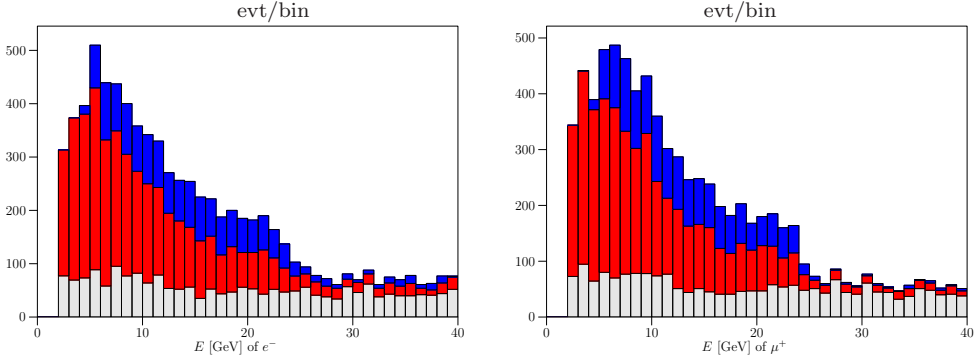

\begin{center}
{\footnotesize evt/bin} \hspace{5cm} {\footnotesize evt/bin}\\
    \includegraphics[width=.4\textwidth]{energiesetc.1}
    \quad\qquad
    \includegraphics[width=.4\textwidth]{energiesetc.2}
    \vspace{2mm}
\caption{\label{fig:lep500} Lepton energy distributions after
applying the cuts: SM background (light
grey), SUSY background (medium grey, red) signal (dark grey, blue)
for the  electrons (left panel) and muons (right panel). We use $\int\mathcal{L} = 1$~ab$^{-1}$. }
\end{center}\vskip -0.5cm
\end{figure}

\section{Sneutrino mass determination}

The most
straightforward way to determine the mass of otherwise invisible sneutrinos is the measurement of the total cross section for the process Eq.~(\ref{eq:signature}). A fit of the collider data
to samples of MC data for varying sneutrino mass (assuming masses of other particles known) allows one to determine the sneutrino mass with
an accuracy of $\sim 0.75$~GeV. A less model-dependent possibility is to use the on-shell kinematics of chargino decay, where the information about particle masses is encoded in the positions of lower and upper edges of the lepton energy spectra. The energy spectra of electrons and muons, including ISR and beamstrahlung,  after cuts are shown in Fig.~\ref{fig:lep500}. The lepton spectrum from SM processes is flat and can easily be subtracted. On the other hand, leptons from the signal and SUSY background exhibit a similar energy dependence. The lower edge of the lepton spectrum is essentially smeared out, mostly due to $p_\perp$ and $\Delta \phi$ cuts which are necessary to kill the photon-induced SM background. The upper edge, however, remains pronounced. If the chargino mass is determined from elsewhere (e.g. the threshold scan), the upper edge allows  one to determine the sneutrino
mass with an accuracy of roughly $1\%$. The most sophisticated method of mass determination by fitting the
entire `experimental' lepton energy distributions to the MC samples generated for varying sneutrino mass allows one to go down with the sneutrino mass error to $0.3$~GeV.

\section{Conclusions}

The chargino pair production process with the subsequent two-body leptonic decay at the ILC
with $\sqrt{s}=500$~GeV has been investigated including all relevant SM and SUSY backgrounds as well as ISR and beamstrahlung. With the event generator {\rm \whizard}, the  full matrix element for production and decay together with off-shell and interference effects were taken into account. We have demonstrated that this process can allow us to determine the sneutrino mass with the accuracy well below $1\%$. In scenarios where sneutrinos are otherwise invisible,
this is the only method of their detection and mass measurement. The expected high experimental precision
at the ILC, however, calls for more precise theoretical methods, so a full NLO calculation of the signal
process is a
possible future improvement of the present study~\cite{nlo}.

\begin{footnotesize}

\section*{Acknowledgments}

Work partially supported by the German BMBF Grant No. 05HA6VFB,
the DFG SFB/TR9 ``Computational Particle Physics'', the Helmholtz-Alliance ``Physics at the Terascale'',  the EU
Network MRTN-CT-2006-035505 ``Tools and Precision Calculations for Physics
Discoveries at Colliders", the EC Programme MTKD-CT-2005-029466
``Particle Physics and Cosmology: the Interface'', as well as by the
Helmholtz-Gemeinschaft Grant No. VH-NG-005 at an early stage of
the work. JK is grateful  to
the Theory Division for the hospitality extended to him at CERN.


\end{footnotesize}


\end{document}